\documentclass{iau} 
\usepackage{graphicx}
\usepackage[square,sort]{natbib}
\usepackage{amssymb}
\usepackage{epstopdf}
\usepackage{natbib}
\usepackage{multicol}
\usepackage{enumitem}
\usepackage{array}

\usepackage[figuresright]{rotating}
\usepackage{lscape}
\usepackage{setspace}
\usepackage{graphics}
\usepackage{epsfig}
\usepackage{multirow} 
\usepackage{bigdelim}
\usepackage{bigstrut}
\usepackage{floatflt}   
\usepackage{wrapfig}
\usepackage{natbib} 
\usepackage{gensymb}

\usepackage[T1]{fontenc}
\usepackage{PTSans}
\def\apj{ApJ}                 
\def\apjl{ApJ}                
\def\apjs{ApJS}               
\def\mnras{MNRAS}             
\def\prd{Phys.~Rev.~D}        
\def\pasa{PASA}               

\usepackage{indentfirst}
\usepackage{pdfpages}

\usepackage{cleveref}
\crefname{section}{§}{§§}
\Crefname{section}{§}{§§}

\title[Strong field tests of gravity with PSR J1141$-$6545] 
{Strong field tests of gravity with PSR J1141$-$6545}

\author[ Venkatraman Krishnan et.al ]   
{ V. Venkatraman Krishnan $^{1,2}$, W. van Straten $^{1,6}$, P. A. Rosado$^{1}$, M. Bailes $^{1,2,3}$, E. F. Keane $^{1,2,4}$, R. Bhat $^{2,5}$, C. Flynn $^{1,2}$}

\affiliation{ $^{1}$ Centre for Astrophysics
 and Supercomputing, Swinburne University of Technology, Mail H30, PO
 Box 218, VIC 3122, Australia. \\ $^{2}$ ARC Centre of Excellence for
 All-sky Astrophysics (CAASTRO). \\ $^{3}$ ARC Centre of Excellence for Gravitational Wave Discovery (OzGrav).\\ $^{4}$ SKA Organisation, Jodrell
 Bank Observatory, SK11 9DL, UK. \\$^{5}$ International Centre
 for Radio Astronomy Research, Curtin University, Bentley, WA 6102,
 Australia.\\ $^{6}$ Institute for Radio Astronomy and Space Research, Auckland University of Technology, Private Bag 92006, Auckland 1142, New Zealand.\\email: {\tt vivekvenkris@gmail.com}}

\pubyear{2017}
\volume{337}  
\jname{Pulsar Astrophysics: The Next Fifty Years}
\editors{P. Weltevrede, B.B.P. Perera, L. Levin Preston \& S. Sanidas, eds.}
\begin{document}

\maketitle

\begin{abstract}
The initial results from timing observations of PSR J1141$-$6545, a relativistic pulsar-white dwarf binary system, are presented. Predictions from the timing baseline hint at the most stringent test of gravity by an asymmetric binary yet. The timing precision has been hindered by the dramatic variations of the pulse profile due to geodetic precession, a pulsar glitch and red timing noise. Methods to overcome such timing irregularities are briefly presented along with preliminary results from the test of the General Theory of Relativity (GR) from this pulsar.
\keywords{pulsars, General Relativity, gravitational waves, strong field tests, pulsar timing}
\end{abstract}

\firstsection 
\section{Introduction}

Pulsars in relativistic binary systems provide the most stringent tests of gravity in the strong field regime to date \citep{KramerEtAl2006,dns,FreireEtAl2012}. While pulsars such as the double neutron star system (B1913+16) and the double pulsar (J0737$-$3039A/B) provide significant tests for the predictions of the GR \citep{KramerEtAl2006,dns}, their gravitational symmetry makes them less sensitive to testing the predictions of alternative theories of gravity such as scalar-tensor theories \citep{DamourEspositoFarese1992,Wex2014}. Such theories are natural extensions of GR that deviate strongly from GR in the strong-field regime, especially in the predictions of multipolar contributions to the gravitational radiation losses in the system. Pulsars in gravitationally asymmetric binaries with white dwarf or black hole companions, such as PSR J1141$-$6545, are better suited systems for testing such theories.\\

PSR J1141-6545 was discovered with the 64-m CSIRO Parkes radio telescope as a part of the Parkes Multibeam Pulsar Survey (PMPS) in 2000 \citep{Kaspietal}. It has a spin period of $\sim$394 ms and is in a binary orbit around a white dwarf with a period of $\sim$ 4.8 hours. Regular timing observations of this pulsar has been carried out since its discovery. Tests of GR using this pulsar with increasing precision, were reported first in 2003 \citep{Bailes2003} and then in 2008 \citep{BhatEtAl2008}. With 2$\times$ increase in the timing baseline since most recently published result, this system is predicted to provide the most stringent test of GR by an asymmetric binary.\\

\section{\label{sc:Observations} Observations and the data reduction pipeline}

The data were taken predominantly using the Parkes telescope with the central beam of the Parkes 20 cm multibeam receiver using several digital signal processors (termed as "backends"). The most recent data after February 2015 has been taken with the newly refurbished UTMOST telescope at 843 MHz \citep{Bailes2017}.\\

The data recording and processing pipelines were built around the \textsc{psrchive}\footnote{psrchive.sourceforge.net} software package \citep{HotanEtAl2004} which in turn used the  \textsc{tempo2}\footnote{https://bitbucket.org/psrsoft/tempo2} \citep{HobbsEtAl2006} timing analysis software to obtain phase predictors to fold the data. The raw data from the telescope was first RFI mitigated, then calibrated for polarisation using the Measurement Equation Template Matching (METM) technique \citep{vanStraten2013} that used observations of the hydra radio galaxy as the flux calibrator and regular observations of PSR J0437$-$4715 over a wide range of parallactic angles as the reference source. The geodetic precession of the pulsar resulting in temporal pulse profile shape variations necessitated the development of a profile evolution model for the generation of dynamic standard templates. A Gaussian component decomposition and evolution algorithm was developed and added to the \textsc{psrchive} package that could model such shape variations and produce analytical (noise-free) standard templates for every observing session (epoch) that are then used to estimate the arrival times. A simultaneous Markov Chain Monte-Carlo (MCMC) fit for the pulsar parameters and a power law model for the red timing noise (as defined in \cite{LentatiEtAl2014}) was done with the \textsc{temponest}369 plugin for the \textsc{tempo2} timing software from which precise pulsar parameters were measured. A more detailed explanation of the data processing techniques will be presented in a forthcoming paper.

\section{Measurements}

Aside from the pulsar spin, astrometric and Keplerian orbital parameters, 3 post-Keplerian parameters were measured : the rate of advance of periastron $\dot{\omega}$, gravitational redshift amplitude $\gamma$ and the orbital period derivative $\dot{\mathrm{P}}_\mathrm{b}$. The intrinsic $\mathrm{\dot{P}}_\mathrm{b}$ due to gravitational radiation losses from the system was obtained by subtracting a kinematic and a Galactic contribution (both values obtained from \citep{BhatEtAl2008}) to the measured $\mathrm{\dot{P}}_\mathrm{b}$. The kinematic contribution is due to an apparent acceleration of the system due to its proper motion, known as the ``Shklovskii'' effect and the Galactic contribution is due to differential rotation in the plane of the Galaxy. The measured Keplerian and post-Keplerian parameters are listed in tables \ref{tab:kep} \& \ref{tab:pk} respectively. The reader is advised to use these values with caution as the results are only preliminary.\\ 

\begin{table}
\caption{Keplerian Parameters for PSR J1141$-$6545. \label{tab:kep}}
\begin{center}

\begin{tabular}{ll}
\hline\hline
Orbital period, $P_b$ (d)\dotfill & 0.19765095945(13) \\
Epoch of period determination (MJD)\dotfill & 51369.9 \\
Epoch of periastron, $T_0$ (MJD)\dotfill & 51369.854549(3) \\
Projected semi-major axis of orbit, $x$ (lt-s)\dotfill & 1.8589232(17) \\
Longitude of periastron, $\omega_0$ (deg)\dotfill & 42.449(5) \\
Orbital eccentricity, $e$\dotfill & 0.1718753(16) \\
\end{tabular}
\vspace{3mm}
\caption{Post-Keplerian Parameters for PSR J1141$-$6545. \label{tab:pk}}

\begin{tabular}{llll}
\hline
Parameter & Measured & GR& Deviation \\
 &  & prediction& from GR \\ \hline
Periastron advance, $\dot{\omega}$ (deg/yr)\dotfill & 5.3099(5) & 5.3104 & $ < 1\sigma$\\
First orbital period derivative (intrinsic), $\mathrm{\dot{P}}_\mathrm{b}$\dotfill & $-$3.93(8)$\times 10^{-13}$ & $-$3.86$\times 10^{-13}$ & $ < 1\sigma$\\
Gravitational Redshift, $\gamma$ (seconds)\dotfill & 7.32(3)$\times 10^{-4}$ & 7.709 $\times 10^{-4}$ & $ \sim 8\sigma$\\

\end{tabular}

\vspace{2mm}
\raggedright 
 \scriptsize{
 {\it Note:}\\
  1. Figures in parentheses are  the nominal 1$\sigma$ \textsc{tempo2} uncertainties in the least-significant digits quoted. \\
  2. Results are preliminary. Use with caution.}
\end{center}

\end{table}

\begin{figure}[b]
\begin{center}
 \includegraphics[scale=0.28]{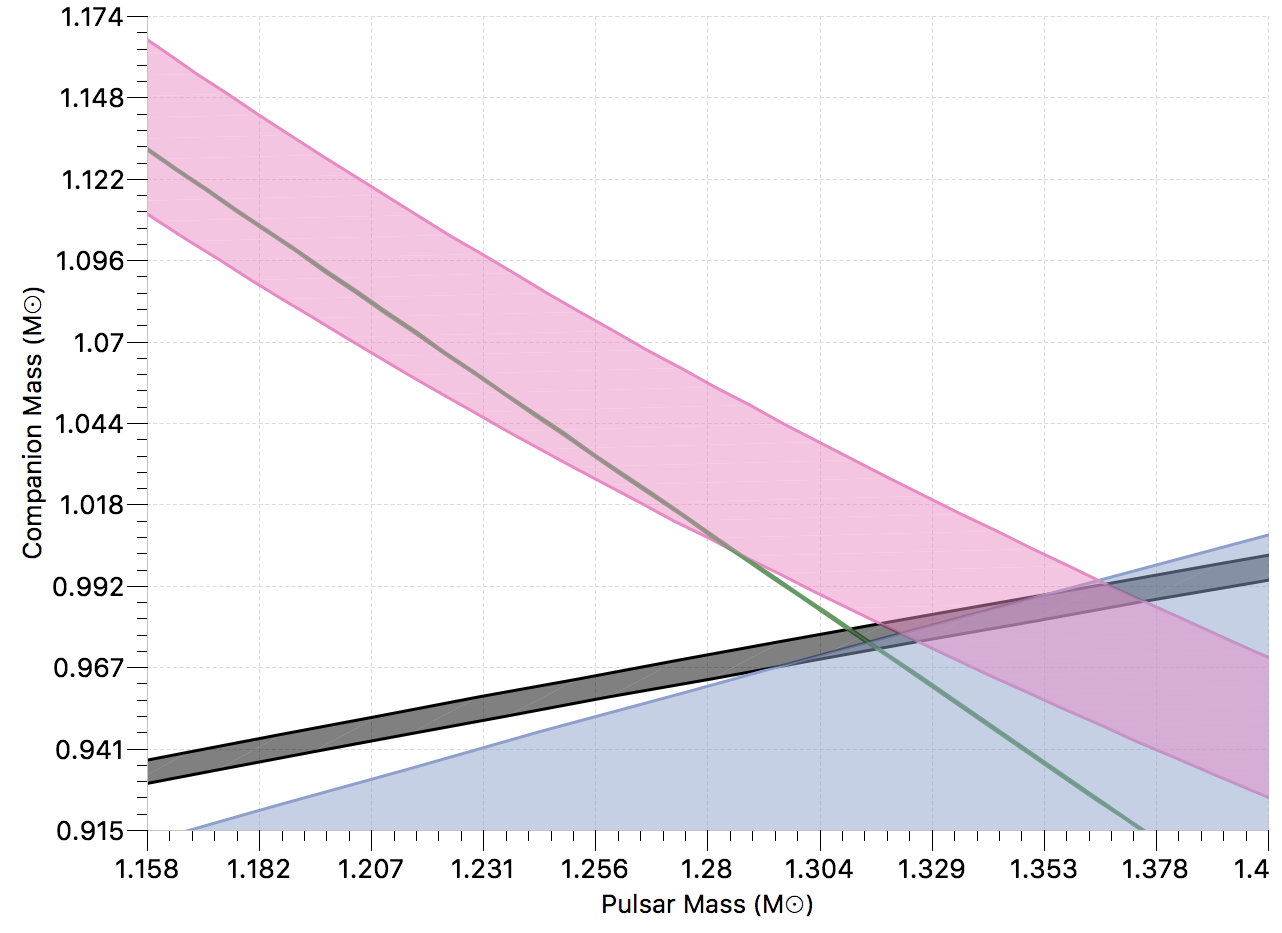} 
 \caption{A preliminary mass-mass diagram plot of the observed post-Keplerian parameters for GR. The pink curve is $\mathrm{\dot{P}}_\mathrm{b}$, green is $\dot{\omega}$, black is $\gamma$ and blue is the mass function. Please see Section \ref{sec:GR} for more details. The shades are 1 $\sigma$ \textsc{tempo2} uncertainties. (The colours can be seen only in the online version).  \label{fig:1}}
   \label{fig1}
\end{center}
\end{figure}

\section{Test of GR}\label{sec:GR}

For any metric theory of gravity such as GR, the post-Keplerian parameters of a binary system can be expressed as a function of the respective masses of the pulsar and the companion. One can then use two of the measured parameters to simultaneously solve for the masses and use any additional parameters as independent tests of that theory of gravity.\\

We carried out a preliminary test of GR with the three measured post-Keplerian parameters from this system. Interestingly, our present results, on first look, point to an inconsistency of the measurements to the predictions of GR. As presented in table \ref{tab:pk}, the predictions of GR for the gravitational redshift parameter $\gamma$, are strikingly different from its measured value. But there are a number of caveats that are to be considered before one can jump to the conclusion that this system has successfully disproved the predictions of GR.\\

Firstly, since the data have undergone rigorous processing techniques, some of which were developed especially for this pulsar, it is important that there are not any residual instrumental effects or data processing artefacts that are systematically affecting the measured values. We performed several tests to identify systematics, including (1) cross-checking the arrival times from simultaneous data streams from multiple backends for a number of observation epochs, (2) cross-checking the arrival times from two different polarisation calibration techniques and (3) cross-checking the confidence intervals obtained from \textsc{tempo2} to the ones from a bootstrap estimator. These checks confirmed the absence of any residual systematic variations in the dataset.\\

Secondly, solving for the component masses using  $\dot{\omega}$ and $\gamma$ and obtaining the corresponding inclination angle via the mass function relation, points to a highly edge-on ($ > 85^{\degree}$) orientation of the binary system. Such an orientation then increases the amplitude of the Shapiro delay, making it a significant measurable (given the RMS value of our timing residuals). Since there is no strong evidence of a Shapiro delay signature in our dataset, such orientation is deemed unlikely, thereby making our measurement of $\gamma$ less credible.\\

Thirdly, $\gamma$ could also be contaminated by other correlated pulsar parameters that are either not yet included in the pulsar model, or are highly co-variant with $\gamma$ thereby driving the best-fit solution to a local minimum. Depending on the timing baseline, the proper motion of the system, orbital period and $\dot{\omega}$, $\gamma$ is co-variant with a number of parameters such as the projected semi-major axis (\textit{x}), secular variations of the semi-major axis ($\dot{x}$) and the relativistic deformation of the orbit ($\delta\theta$), to name a few. We are currently investigating such correlations and the results will be presented in a forthcoming paper.

\end{document}